\documentclass[11pt]{article}
\usepackage{moriond,epsfig,graphicx}

\def\Journal#1#2#3#4{{#1} {\bf #2}, #3 (#4)}

\def\NIMA{{\em Nucl. Instrum. Methods} A}
\def\NPB{{\em Nucl. Phys.} B}
\def\PLB{{\em Phys. Lett.}  B}

\def\PRD{{\em Phys. Rev.} D}

\def\be{\begin{equation}}
\def\ee{\end{equation}}
\def\bea{\begin{eqnarray}}
\def\eea{\end{eqnarray}}

\makeatletter
  \newcommand\figcaption{\def\@captype{figure}\caption}
  \newcommand\tabcaption{\def\@captype{table}\caption}
\makeatother

\def \bm {$B$-meson}
\def \babar {BaBar}
\def \vub {\ensuremath{| V_{ub} |}}
\def \bsg {\ensuremath{B\to X_s \gamma}}
\def \mupi {\ensuremath{\mu_\pi^2}}
\def \mb {\ensuremath{m_b}}
\def \BB {\ensuremath{B\bar{B}}}

\def \bz {\ensuremath{B^0}}
\def \bp {\ensuremath{B^+}}
\def \qsq {\ensuremath{q^2}}

\def \mESDef{\ensuremath{m_{\rm ES} = \sqrt{s/4-|\vec p^{\,*}_B|^2}}}
\def\DeltaEDef{\ensuremath{\Delta E = E^*_B-\sqrt{s}/2}}
\newcommand {\errs} [2] {\ensuremath{^{+#1}_{-#2}}}

\begin{document}
%\vspace*{4cm}
\title{SEMILEPTONIC AND ELECTROWEAK PENGUIN RESULTS FROM BABAR}

\author{ JOHN WALSH~\footnote{john.walsh@pi.infn.it} \\
(on behalf of the BaBar Collaboration)}

\address{INFN, Sezione di Pisa, Largo Pontecorvo 3, \\
56127 Pisa, Italia}

\maketitle\abstracts{We report recent results from the \babar\
  experiment on semileptonic charmless \bm\ decays and electroweak
  penguin processes. Semileptonic charmless decays are used to
  determine \vub\ and the exclusive modes considered here also begin
  to constrain QCD-lattice form factor calculations. Radiative
  penguin decays are both sensitive to physics beyond the Standard
  Model and can be used to extract Heavy Quark parameters related to
  the $b$-quark mass and its motion inside the hadron.
}

\section{Introduction}

An important goal of the study of semileptonic charmless \bm\ decays
is the measurement of \vub, which essentially measures one side of the
Unitarity Triangle of the CKM matrix. Both inclusive and exclusive
analyses have been used to measure \vub, here we report on recent
results using exclusive decays. The measurement of exclusive branching
fractions is also useful for distinguishing among theoretical
calculations of the form factors, as we shall see.

The the radiative decay \bsg\ is studied as a probe of New
Physics. Since this decay occurs at the one-loop level, the branching
fraction is sensitive to models with additional heavy particles that
can participate in the loop. In contrast, the
shape of the photon energy spectrum is quite insensitive to
contributions from New Physics, but it is rather sensitive to two
important parameters of Heavy Quark (HQ) theory: the $b$-quark mass \mb\
and the quantity \mupi, which is related to the Fermi motion of the
$b$-quark inside the hadron.

In this report, we present recent results from the \babar\
experiment~\cite{bib:nim} on semileptonic charmless \bm\ decays to
exclusive states and on the $B\to X_s\gamma$ process. Beyond providing
information on the CKM matrix and probing the possibility of New
Physics, these analyses also provide insight into \bm\ decay dynamics
and QCD. All results presented herein are preliminary.

\section{Exclusive semileptonic charmless $B$ decays}
Semileptonic charmless \bm\ decays to exclusive final states can be
used to measure \vub\ by exploiting the dependence of the branching
fraction on the CKM matrix element. In the case of
$\bz\to\pi^-\ell^+\nu$, we have:
\begin{equation}
\frac{d\Gamma(\bz\to\pi^-\ell^+\nu)}{dq^2} =
    \frac{G_F^2}{24\pi^3}|V_{ub}|^2 p_\pi^3 |f_+(\qsq)|^2
\end{equation}
Here $G_F$ is the Fermi coupling constant, $p_\pi$ is the pion
momentum in the center-of-mass frame, $q$ is the invariant mass of the
lepton-neutrino pair and $f_+(\qsq)$ is the form
factor, which is calculated theoretically. The goal is to measure the
branching fraction in bins of \qsq, which allows one to distinguish among
form factor calculations as well as extract the value of \vub. 

Experimentally, the branching fractions of exclusive $b\to u$ decays
are small and backgrounds from $b\to c$ transitions are substantial.
\babar\ has used two different methods for overcoming the experimental
difficulties: 1) an ``untagged'' analysis, based on 83 million \BB\
pairs, where a premium is placed on high quality neutrino
reconstruction using the missing momentum in the event; and 2) a
``tagged'' analysis (232 million \BB\ pairs for the
$\bz\to\pi^-\ell^+\nu$ state, 88 million \BB\ pairs for
$\bp\to\pi^0\ell^+\nu$ state), where backgrounds are reduced by
requiring the other \bm\ in the event be ``tagged'' via a
$D^{(*)}\ell\nu$ decay.

\begin{figure}
\begin{center}
\epsfig{figure=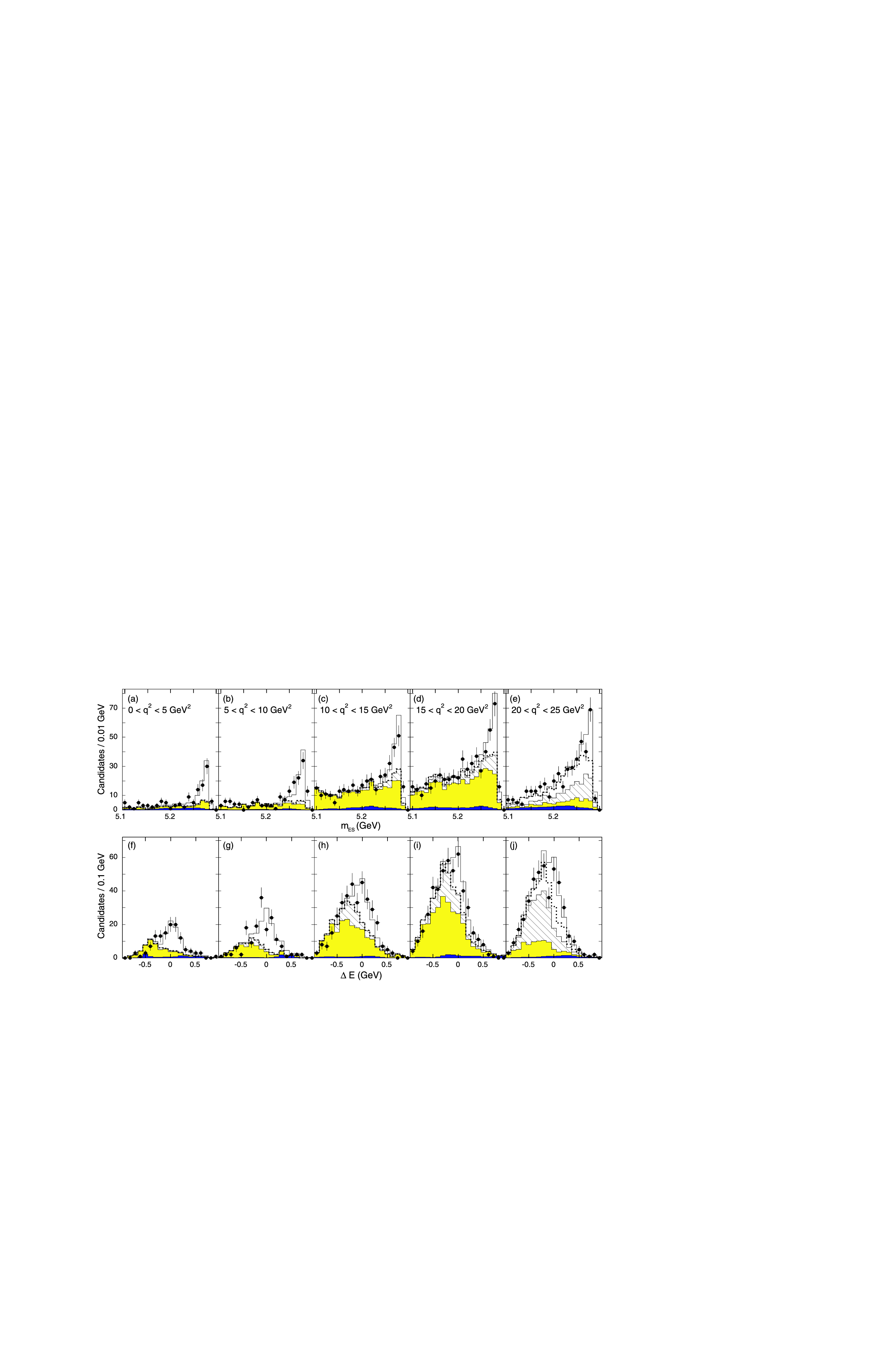,width=0.75\textwidth}
\caption{Projected $m_{ES}$ (top) and $\Delta E$ (bottom) distributions
in five bins of \qsq\ for the $B\to\pi\ell\nu$ modes. The data is
represented by points with (statistical) error bars, while the
simulation is shown as a histogram. The simulated components are:
signal (white), combinatoric signal (dashed), crossfeed (hatched),
$b\to c\ell\nu$ (light shaded) and continuum (dark shaded). 
}
\label{fig:mes}
\end{center}
\end{figure}

The untagged analysis relies on good neutrino reconstruction to
perform its measurement of the branching fractions of $B\to\pi\ell\nu$
and $B\to\rho\ell\nu$. The neutrino momentum is inferred from the
event missing momentum and strict requirements are placed to ensure
good neutrino reconstruction. For example, the event missing mass is
required to be compatible with zero: since its resolution broadens
linearly with missing energy, we require
$|m^2_{\mathrm{miss}}/2E{\mathrm{miss}}|<0.4$ GeV. The variables used
to distinguish signal from background are \mESDef\
and \DeltaEDef, where $\sqrt{s}$ is the total energy in the
$\Upsilon(4S)$ center-of-mass frame. Figure~\ref{fig:mes} shows the
distribution of these two variables for the $B\to\pi\ell\nu$ modes in
five bins of \qsq. 
Branching fraction measurements from both the tagged and untagged
analyses are reported in Table~\ref{tab:bf}. 
%
% the following does some latex tricks to put a table next to a figure
% 
\begin{table}
\begin{minipage}[b]{0.53\textwidth}
\centering
\caption{Branching fractions to charmless semileptonic exclusive
  states from the tagged (labeled ``T'') and untagged (``U'')
  analyses. The uncertainties shown are statistical, systematic and
  (for the untagged analysis) due to form factor uncertainties. The
  untagged analysis assumes the isposin relations:
  $B(\bz\to\pi^-(\rho^-)\ell^+\nu) =
  2B(\bp\to\pi^0(\rho^0)\ell^+\nu)$}
\label{tab:bf}
\centering
\resizebox{.9\textwidth}{!} {
\begin{tabular}{|c|c|} \hline
Mode (Technique) & Branching Fraction ($10^{-4}$)\\ \hline
$\bz\to\pi^-\ell^+\nu $ (T) & $1.03 \pm 0.25 \pm 0.13 $ \\
$\bp\to\pi^0\ell^+\nu $ (T) & $1.80 \pm 0.37 \pm 0.23 $ \\
$\bz\to\pi^-\ell^+\nu $ (U) & $1.38 \pm 0.10 \pm 0.18 \pm 0.08 $ \\
$\bz\to\rho^-\ell^+\nu$ (U) & $2.14 \pm 0.21 \pm 0.53 \pm 0.28 $ \\
\hline
\end{tabular} }
\end{minipage} \hfill
\begin{minipage}[b]{0.46\textwidth}
\centering
\includegraphics[width=0.95\textwidth]{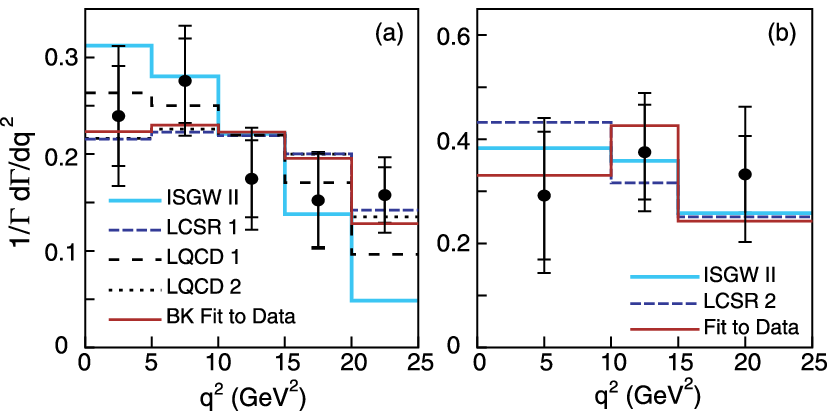}
\figcaption{Comparison of the differential decay rate from the untagged
  analysis for $B\to\pi\ell\nu$ (left) and $B\to\rho\ell\nu$ (right)
  together with several theoretical form factor calculations.}
\label{fig:ff}
\end{minipage}
\end{table} 

The statistics of the untagged sample permits the study of the \qsq\
dependence of the branching fraction and an investigation of several
form factor calculations. Figure~\ref{fig:ff} shows the differential
decay rates along with the predictions of four theoretical
calculations: LCSR1~\cite{bib:lcsr1}, LQCD1~\cite{bib:lqcd1},
LQCD2~\cite{bib:lqcd2} and ISGW II~\cite{bib:isgw}. The $\chi^2$
probabilities are good ($\sim 50$\%) for the first three calculations, while
it is marginal (3\%) for the ISGW II prediction. We extract the value
of \vub\ using the $B\to\pi\ell\nu$ data and the LQCD2 calculation
over the full \qsq\ range $0-25$ GeV$^2$. The BK
parametrization~\cite{bib:bk} is used to extrapolate the LQCD2 form
factor calculation to low \qsq. We obtain 
$\vub = (3.82 \pm 0.14 \pm 0.24 \pm 0.11^{+0.88}_{-0.52})\times
10^{-3}$, where the uncertainties are due to statistics, systematics,
form factor shape and form factor normalization, respectively.

\section{\bsg}

\babar\ has performed two analyses of the \bsg\ channel: a fully
inclusive measurement, where no requirements are made on the hadronic
state ($X_s$) and a semi-inclusive analysis, which aims to reconstruct
a large part of the total \bsg\ rate by summing many exclusively
reconstructed modes. The two approaches are complementary: the fully
inclusive method requires a lepton tag to reduce continuum background,
but nevertheless suffers from significant backgrounds from \BB\
events.  The semi-inclusive analysis, which sums 38 exclusive decay
modes, has the advantage of reduced backgrounds due to the kinematic
handles provided by fully reconstructed $B$ candidates. This analysis
however, has a significant systematic uncertainty due to the {\em
missing fraction}, the part of the \bsg\ rate that it does not
reconstruct.  Both of these analysis are based on approximately 89
million \BB\ pairs.

%Both analyses require a well-reconstructed isolated photon with energy
%above 1.9 GeV. Photons coming from \piz\ and \et\ decay are explicitly
%vetoed. The fully inclusive analysis greatly reduces the significant
%background from continuum events by requiring a lepton tag in the
%event. This lepton is produced by the semileptonic decay of the other
%\bm\ in the event. Event shape criteria further remove some of the
%continuum background. Data taken off-resonance is used to subtract the
%remaining continuum background. Backgrounds from \BB\ events are
%estimated using a Monte Carlo simulation; for the most important
%backgrounds the simulation is checked and corrected using data control
%samples. 

\begin{figure}
\begin{center}
\epsfig{figure=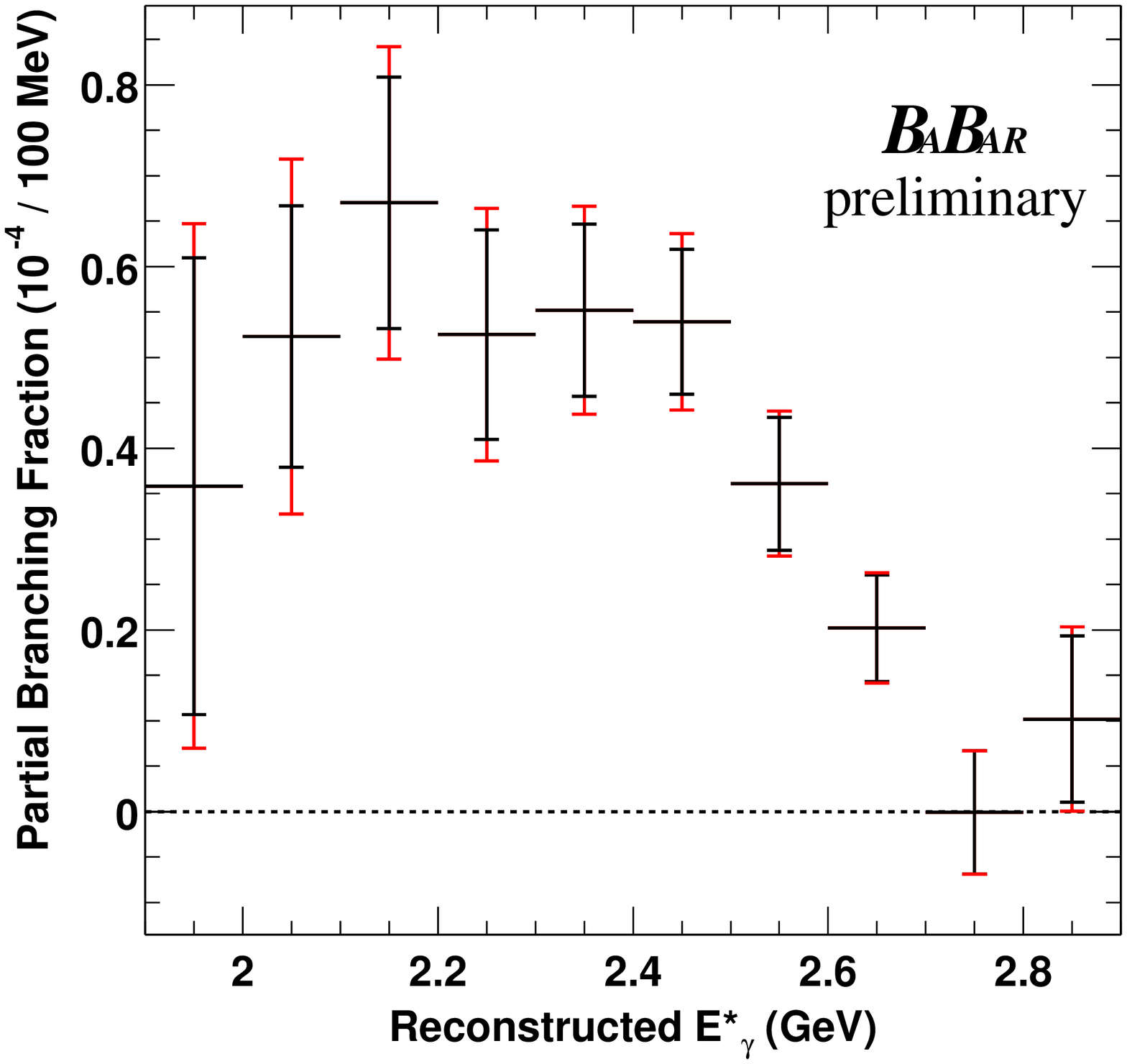,width=0.35\textwidth}
\epsfig{figure=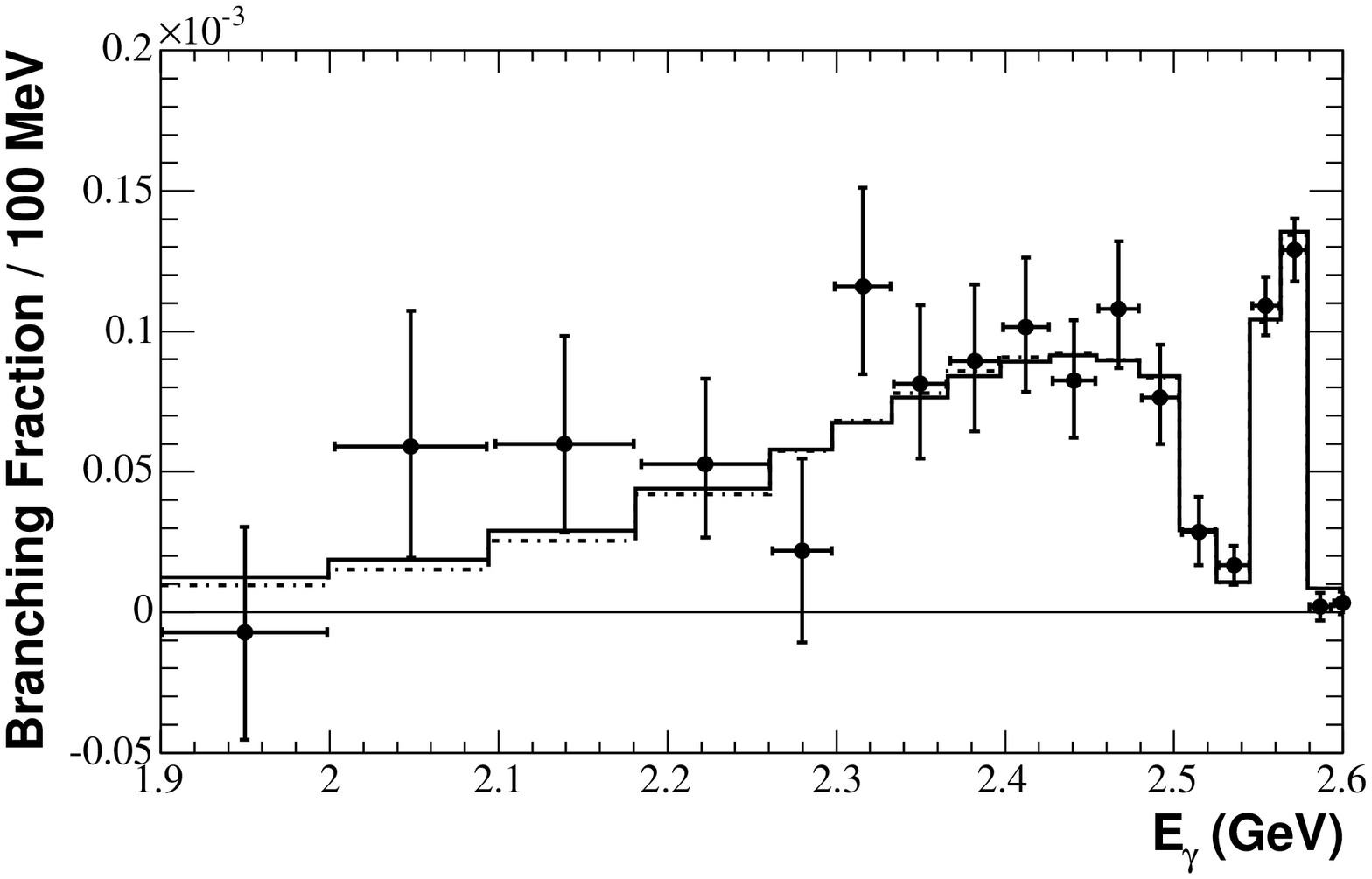,width=0.48\textwidth}
\caption{The photon energy spectrum, background-subtracted and
  efficiency-corrected, for the fully inclusive analysis (left) and
  the semi-inclusive analysis (right). The fully inclusive spectrum is
  measured in the $\Upsilon(4S)$ rest frame, while the semi-inclusive
  spectrum is measured in the \bm\ rest frame.  The peak due to $B\to
  K^{*}\gamma$ decays at high photon energy of the semi-inclusive
  spectrum is visible due to the good resolution (see text).  }
\label{fig:bsg_spectra}
\end{center}
\end{figure}

Figure~\ref{fig:bsg_spectra} shows the resulting photon energy spectra
for the two analyses. The semi-inclusive analysis has better photon
energy resolution for two reasons: 1) the energy is measured in the
\bm\ rest frame and 2) the photon energy is actually inferred from the
hadronic invariant mass, which has quite good resolution.
We present in Table~\ref{tab:fi_moments} the energy moments of the
photon spectrum calculated above a certain energy threshold, measured
in the \bm\ rest frame. A correction is applied to the fully-inclusive
values to bring them into this frame. These moments may be directly
compared to theoretical calculations to give information on HQ
parameters.  A fit to the semi-inclusive spectrum was performed to
extract the HQ parameters \mb\ and \mupi. Two theoretical schemes were
used to perform the fits: the kinetic scheme~\cite{bib:kinetic}, which
gives:
\begin{equation}
\mb = 4.69 \errs{0.05}{0.04}\ \mathrm{GeV} \quad \mathrm{and} \quad 
\mupi = 0.30\errs{0.07}{0.05}\ \mathrm{GeV}^2;
\end{equation}
and the shape function scheme~\cite{bib:sf}, which yields:
\begin{equation}
\mb = 4.65 \pm 0.04 \ \mathrm{GeV} \quad \mathrm{and} \quad 
\mupi = 0.19\errs{0.06}{0.05}\ \mathrm{GeV}^2;
\end{equation}
where the errors are the sum of statistical and systematic, but do not
include theoretical uncertainties. We note that the parameter \mupi\
is not defined the same way in the two schemes. The spectrum fit also
yields the total inclusive branching fraction down to $E_\gamma>1.6$
GeV. Averaging the results from the two theoretical schemes gives: 
$B(b\to s\gamma, E_\gamma > 1.6\ \mathrm{GeV})= (3.38 \pm
0.19^{+0.64+0.07}_{-0.41-0.08})\times 10^{-4}$. We note that the
branching fraction result is compatible with the Standard Model
calculation~\cite{bib:bsg_sm} and with the experimental world
average~\cite{bib:hfag}.
\begin{table}
\label{tab:fi_moments}
\caption{Photon energy spectrum moments and partial branching
fractions (PBF) from $B\to X_s\gamma$. The first four rows are from
the fully inclusive analysis, the remaining rows are from the
semi-inclusive analysis. Values are given for the \bm\ rest frame,
including the minimum photon energy. The uncertainties shown are
statistical, systematic and model-dependent (for the fully inclusive
analysis), respectively. }
\begin{center}
\renewcommand{\arraystretch}{1.25}
\resizebox{\textwidth}{!} {
\begin{tabular}{|c|cccc|} \hline
Min $E_\gamma$ & PBF ($10^{-4}$)                     & 1st Moment (GeV) & 2nd Moment (GeV$^2$) & 3rd Moment (GeV$^3$) \\ \hline
1.9            & $3.67 \pm 0.29 \pm 0.34 \pm 0.29 $ & $2.288 \pm 0.025 \pm 0.017 \pm 0.012$ &- &-\\
2.0            & $3.41 \pm 0.27 \pm 0.29 \pm 0.23 $ & $2.316 \pm 0.016 \pm 0.010 \pm 0.012$ &- &-\\
2.1            & $2.97 \pm 0.24 \pm 0.25 \pm 0.17 $ & $2.355 \pm 0.014 \pm 0.007 \pm 0.010$ &- &-\\
2.2            & $2.42 \pm 0.21 \pm 0.20 \pm 0.13 $ & $2.407 \pm 0.012 \pm 0.005 \pm 0.008$ &- &-\\ \hline
1.897 &- & $2.321 \pm 0.044 \errs{0.037}{0.026}$ & $0.0253 \pm 0.0116 \errs{0.0049}{0.0042} $ & $0.0006 \pm 0.0085 \errs{0.0041}{0.0032} $ \\
1.999 &- & $2.314 \pm 0.025 \errs{0.026}{0.027}$ & $0.0273 \pm 0.0039 \errs{0.0042}{0.0037} $ & $0.0009 \pm 0.0036 \errs{0.0036}{0.0022} $ \\
2.094 &- & $2.357 \pm 0.018 \errs{0.014}{0.016} $ & $0.0183 \pm 0.0023 \errs{0.0021}{0.0012} $ & $0.0005 \pm 0.0017 \errs{0.0016}{0.0009} $ \\
2.181 &- & $2.396 \pm 0.013 \errs{0.008}{0.004}$ & $0.0115 \pm 0.0014 \errs{0.0010}{0.0006} $ & $0.0001 \pm 0.0008 \errs{0.0006}{0.0003} $ \\
2.261 &- & $2.425 \pm 0.009 \errs{0.004}{0.002}$ & $ 0.0075 \pm 0.0007 \errs{0.0003}{0.0002} $ &  $-0.0001 \pm 0.0003 \errs{0.0002}{0.0001} $ \\
\hline
\end{tabular} } %end of resizebox
\end{center}
\end{table}

\section*{Acknowledgments}

A heartfelt thanks the organizers for a stimulating and enjoyable
conference. Thanks also go to my \babar\ colleagues for their assistance
in preparing these results.

\end{document}